\documentclass[pdflatex,sn-mathphys-num,iicol]{sn-jnl}

\usepackage{graphicx}%
\usepackage{multirow}%
\usepackage{amsmath,amssymb,amsfonts}%
\usepackage{amsthm}%
\usepackage{mathrsfs}%
\usepackage[title]{appendix}%
\usepackage[dvipsnames]{xcolor}% dvipsnames needed for ForestGreen/Maroon in the author-colour macros below
\usepackage{textcomp}%
\usepackage{manyfoot}%
\usepackage{booktabs}%
\usepackage{algorithm}%
\usepackage{algorithmicx}%
\usepackage{algpseudocode}%
\usepackage{listings}%
\theoremstyle{thmstyleone}%
\usepackage{tikz}
\usepackage{hyperref}
\usepackage{enumitem}
\usepackage{listings}
\usepackage{tabularx}
\usepackage{caption}
\usetikzlibrary{
  arrows.meta,
  positioning,
  shapes.geometric,
  decorations.pathreplacing,
  calc,
  fit,
  backgrounds
}
\usepackage{bbold}

\newcolumntype{Y}{>{\centering\arraybackslash}X} % Centered text wrapping

\theoremstyle{thmstyletwo}%

\theoremstyle{thmstylethree}%

 \newcounter{mcomments}

 \newcounter{Dcomments}

\newlist{outline}{enumerate}{3}
\setlist[outline,1]{label=\Roman*., leftmargin=2cm, labelsep=0.5cm}
\setlist[outline,2]{label=\Alph*., leftmargin=1cm, labelsep=0.5cm}
\setlist[outline,3]{label=\arabic*., leftmargin=1cm, labelsep=0.5cm}

\newif\ifauthorcolours
\authorcolourstrue

\DeclareTextCommand{\k}{OT1}[1]{#1}

\begin{document}

\title[Towards Optimal Routing Strategies for qLDPC Codes on Silicon Spin Qubits]{Towards Optimal Routing Strategies for qLDPC codes on Silicon Spin Qubits}

\author*[1]{\fnm{Merlin} \sur{Incerti-Medici}}\email{merlin@mothquantum.com}

\author[2]{\fnm{Minh} \sur{T. P. Nguyen}}

\author[2]{\fnm{Stefano} \sur{Bosco}}

\author[1]{\fnm{James} \sur{R. Wootton}}

\author*[1]{\fnm{Daniel} \sur{Bultrini}}\email{daniel@mothquantum.com}

\affil[1]{\orgname{Moth}, \orgaddress{\street{Schorenweg 44b}, \city{Arlesheim}, \postcode{4144}, \state{Basel-Land}, \country{Switzerland}}}

\affil[2]{\orgdiv{QuTech and Kavli Institute of Nanoscience}, \orgname{Delft University of Technology}, \orgaddress{\street{Lorentzweg 1}, \city{Delft}, \postcode{2628CJ}, \state{South Holland}, \country{The Netherlands}}}

\abstract{One of the major challenges of implementing high-rate quantum error correction codes, like qLDPC codes, is realizing the required long-range interactions. Spin qubit platforms can provide this connectivity by shuttling qubits next to each other whenever they need to interact. When doing this, the performance of the code depends on the shuttling strategy. We study memory experiments of the [[144,12,12]] bivariate bicycle (Gross) code on two shuttling grids for future spin qubit devices: a square grid and a square grid with diagonals. We compare three shuttling strategies, structured and random initial layouts, and a range of shuttling noise models spanning best- and worst-case scenarios. The best strategy minimizes the longest distance travelled rather than the total distance. The initial layout has only a weak effect, whereas adding diagonals to the grid reduces the logical error rate by up to two orders of magnitude which may have implications for future hardware design.}




\maketitle

\section{Introduction}\label{Introduction}

The successful deployment of quantum error correction (QEC) to preserve information in the presence of noise is foundational to achieving large scale fault tolerant quantum computers (FTQC). In recent years, QEC has been explored on a range of platforms, including superconducting qubits \cite{googlequantumaiandcollaboratorsQuantumErrorCorrection2025, wangDemonstrationLowoverheadQuantum2026}, trapped ions \cite{thamBreakevenDemonstrationQuantum2026}, neutral atoms \cite{bluvsteinLogicalQuantumProcessor2024}, and semiconductor spin qubits \cite{takedaQuantumErrorCorrection2022, undsethWeightfourParityChecks2026, teamDigitallyControlledSilicon2026}. The two dimensional surface code has been the QEC code of choice due to its high threshold and planar nearest neighbour connectivity requirements \cite{fowlerSurfaceCodesPractical2012}. Although surface codes have been optimized and have been experimentally successful \cite{googlequantumaiandcollaboratorsQuantumErrorCorrection2025}, their asymptotic encoding rate is zero, meaning that the number of physical qubits required to encode a single logical qubit with sufficiently low error rates gets prohibitively large. High-rate quantum low-density parity-check (qLDPC) codes \cite{breuckmannQuantumLowDensityParityCheck2021, tillichQuantumLDPCCodes2014, panteleevQuantumLDPCCodes2022, breuckmannBalancedProductQuantum2021} have been proposed as an attractive alternative. Their appeal is a trade-off: encoding a larger number of logical qubits in a single code block while maintaining a large distance and a competitive threshold at the expense of requiring parity check of (slightly) higher weight that act on qubits that are non-local in any planar layout \cite{breuckmannQuantumLowDensityParityCheck2021, baspinQuantifyingNonlocalityHow2022}. The family of qLDPC codes provides a useful class of codes within which this trade-off can be explored. For this paper, we focus on the Bivariate Bicycle (BB) code with parameters [[144, 12, 12]], known as the Gross code \cite{bravyiHighthresholdLowoverheadFaulttolerant2024}.

BB codes illustrate the trade-off of higher-weights and non-locality very well: each qubit of the Gross code has to interact with six data qubits (compared to four in the surface code). The resulting Tanner graph is a non-planar graph of degree six. While the Tanner graph of a BB code is the union of two edge-disjoint planar subgraphs, the full graph cannot be made planar and in any physical realization of the full graph, some qubits that have to interact are far apart. 

In platforms where qubits can be physically moved, shuttling is a viable way to enable far-apart qubits to interact. Shuttling can be done for a variety of qubit platforms, making it a relevant method beyond one singular approach to FTQC. The platforms that allow shuttling include neutral atoms which can be rearranged with optical tweezers \cite{bluvsteinLogicalQuantumProcessor2024, xuConstantoverheadFaulttolerantQuantum2024, cainShorsAlgorithmPossible2026}, trapped ions which are shuttled between the zones of segmented traps \cite{pinoDemonstrationTrappedionQuantum2021, durandauShuttlingBidimensionalSegmented2026}, and semiconductor spin qubits. We focus specifically on semiconductor spin qubits, which encode quantum information in the spin states of electrons or holes confined in quantum dots \cite{burkardSemiconductorSpinQubits2023}. A growing number of coherent spin shuttling experiments in silicon and germanium \cite{noiriShuttlingbasedTwoqubitLogic2022,struckSpinEPRpairSeparationConveyormode2024,xueSiSiGeQuBUS2024,vanriggelen-doelmanCoherentSpinQubit2024,desmetHighfidelitySinglespinShuttling2025,fosterDephasingErrorDynamics2024,matsumotoTwoqubitLogicTeleportation2026} have shown that these qubits can be transported on the chip. In addition, architectures for spin qubits that utilize this shuttling have been explored \cite{caiLoopedPipelinesEnabling2023, patomakiPipelineQuantumProcessor2024, siegelQuantumSnakesPlane2026}. This shuttling capability suggests that it might be possible to run QEC codes with non-local connections, that is with parity checks that couple qubits located far apart. Schemes to do this effectively have recently been explored in \cite{chadwickCAbLECAREfficientlyScheduling2026, siegel2026nestultrahighrateqldpccodes}.

In this work, we systematically evaluate the performance of the Gross code on a shuttling based spin qubit architecture. To do so, we develop several algorithms to determine shuttling routes for the qubits to execute a BB-code. We then compare the shuttling routes that these algorithms produce for their efficiency. In addition to comparing these different methods of computing shuttling routes, we compare two distinct shuttling grids. Finally, we compare different starting positions for the qubits before running a syndrome extraction circuit. Our key findings are:

\begin{enumerate}
    \item Reducing the largest shuttling distance is more effective at preventing logical errors than reducing the overall amount of shuttling.
    
    \item Increasing the connectivity of the shuttling grid significantly improves the performance of the error correction code making this worthwhile to pursue in practical architectures.
    
    \item The starting positions of the qubits have only a minor effect on the performance of the error correction code.
\end{enumerate}

Our analysis provides some immediate takeaways for shuttling capabilities and for building hardware architectures. The maximal necessary shuttling distance should be kept low whenever possible and building a shuttling grid with higher connectivity is worth the effort, as long as this does not cause an enormous amount of additional noise. Furthermore, the low effect of the starting position allows for greater flexibility when preparing the qubits for error correction.

The code for our simulation is written in such a way that various inputs can easily be replaced, including the error correction code (provided it is a BB code), the shuttling grid, the shuttling strategy, or the noise model. The code can thus serve as a basis for future tests to optimize shuttling for error correction in spin qubits.

The rest of the paper is organized as follows: In Section \ref{sec2} we present our results: we explain the basics of shuttling strategies and the way we built the strategies we tested, we present the outcomes of the simulations, and we discuss the most important findings. In Section \ref{methods}, we discuss the key assumptions we make about future spin qubit hardware, the shuttling grids we compare, the starting positions we give the qubits, the noise models we use, and the algorithms that compute our shuttling routes. Section \ref{Discussion} is the discussion of our results. Appendix \ref{app:NoisyCircuit} goes into the process how we add noise to the ideal syndrome extraction circuit of the memory experiment. The full results of our simulations are presented in Appendix \ref{datasweep}. Appendix \ref{app:algorithms} contains the algorithms and figures.

\section{Results}\label{sec2}

QEC codes require the interaction of many physical data and check qubits. In the setting considered in this paper, these qubits are distributed over a finite physical grid (see Fig.~\ref{fig:grids} for the two grids considered in this work) and must be moved near each other to undergo a two-qubit interaction. To do this one must determine a routing logic to determine all the qubits' movement. We refer to these as shuttling strategies.

Since there are many potential variances in existing and potential future spin qubit hardware, this work compares different shuttling strategies for a large space of parameters and searches for error behaviour that is robust under parameter change. Specifically, we test our shuttling strategies under combinations of different shuttling dephasing noise models (linear, exponential and Gaussian),  initial layouts, shuttling grids, ratios between the shuttling to CNOT error rates, and idling to shuttling ratios. These are described in Sec.~\ref{subsec:NoiseModels}.

Our analysis provides strong evidence that the implementation of an optimal shuttling strategy is important to reduce logical error rates. This effect is amplified when the shuttling error rate increases relative to the two-qubit gate error rate or if a Gaussian noise model is used instead of a linear or exponential noise model.\footnote{In practice, a Gaussian noise model corresponds for example to the standard quasi-static dephasing of a qubit that is either stationary or moves slowly enough to experience quasi-static noise.} We also find that the initial coordinates of the qubits have a lesser effect on the resulting logical error rate, while a higher connectivity of the shuttling grid can yield greatly improved results. Overall, these findings point towards a general heuristic that logical error rates are suppressed if we minimize the largest shuttling distance rather than the total sum of shuttling distances and that higher connectivity is a valuable feature to work on when designing hardware architecture.

\subsection{Shuttling Strategies} \label{subsec:QubitShuttling}

\subsubsection*{Basics}

\begin{figure*}[htpb]
	\centering
    \includegraphics[scale=1]{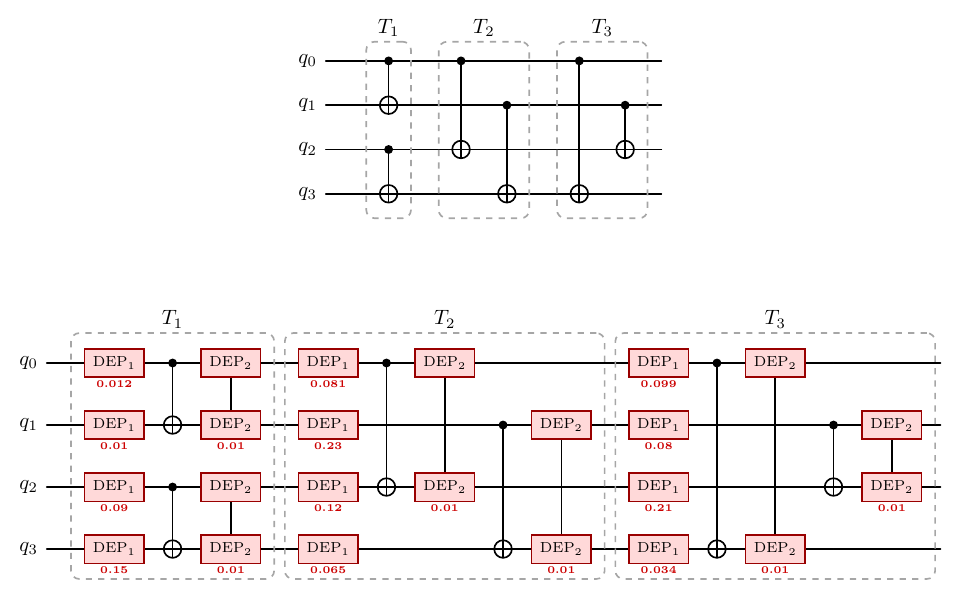}
    \caption{On the top we see a noise-free circuit, split into three ticks $T_1$, $T_2$, $T_3$. For each tick, we determine shuttling routes that need to be done before the tick, which causes shuttling and idling noise, depicted in the bottom circuit as $\text{DEP}_1$ noise channels in red as single qubit noise channels. The red noise channels labeled $\text{DEP}_2$ represent a two qubit noise channel caused by 2 qubit gates, which are applied after the gate.}
    \label{fig:adding_noise}
\end{figure*}

Consider a quantum circuit $C$ that can be split into a sequence of ticks $T_0, T_1, \dots$ where a tick is defined by a slice of a circuit such that all gate operations within that slice  can be performed in parallel on a given machine. We add shuttling to this circuit as follows: Suppose we know the position of every qubit at the end of tick $T_{i-1}$. For every qubit $q$ we look what it has to do in tick $T_i$. If $q$ does not have to interact with any other qubit during $T_i$, we move $q$ to a spot $target(q)$ with no other qubit where its operation during $T_i$ can be executed. If $q$ has to do a 2-qubit gate with another qubit, say $q'$, we move both $q$ and $q'$ to the same spot $target(q) = target(q')$, where said 2-qubit gate can be executed. The positions we move the qubits to are their positions after tick $T_i$. We then repeat this for tick $T_{i+1}$. We refer to the shuttling that happens between $T_{i-1}$ and $T_i$ as one {\it shuttling phase}. The concept of ticks and how we add noise to each tick is illustrated in Fig.~\ref{fig:adding_noise}. 

Different shuttling strategies are distinct methods to determine the spots $target(\cdot)$ in a shuttling phase. Our goal is to find strategies that are both fast to compute  and keep the logical error rate low. While shuttling routes can be computed before running the quantum computer, it is still useful to be able to compute them fast, since -- even for a fixed circuit like syndrome extraction of a chosen error correction code -- the optimal route may have to be recalibrated on a regular basis. In principle, we want to reduce the logical error rate by optimizing across all shuttling phases together. Doing so might require doing suboptimal shuttling during individual shuttling phases. However, optimizing over all shuttling phases has a substantially higher complexity than optimizing each shuttling phase individually, which is why we look at each shuttling phase in isolation.

\subsubsection*{Heuristic rules}

Suppose we know the current location of all qubits and we are given a tick $T$. How do we determine the locations $target(\cdot)$ that every qubit has to go to for tick $T$? We will build different strategies for finding $target$ that are based on the following two rules:

\begin{enumerate}
    \item[Rule 1] Find $target(\cdot)$ so that it minimizes the largest shuttling distance during the shuttling phase.

    \item[Rule 2] Find $target(\cdot)$ so that it minimizes the total sum of all shuttling distances during the shuttling phase.
\end{enumerate}

As we demonstrate with our simulations (see Fig.\,\ref{fig:comparison_main}), rule 1 provides a significant improvement in performance. Aside from lookahead, which is left to future work, the most promising alternative rule to the two above is to minimize the largest idling time in each shuttling phase. This is particularly important given the fact that idling causes more noise per time than shuttling in some devices \cite{beerConveyormodeElectronShuttling2026}. However, minimizing the largest shuttling distance also contributes a lot to minimizing the largest idling time while being much more computationally accessible. For this reason, we focus on the two rules above and the strategies that can be built from them. One could also consider minimizing other statistical properties like the median, but after our simulations it was clear that rule 1 was the most important.

The three shuttling strategies that we will compare are built from the two rules above as follows:

\begin{enumerate}
    \item[ ] {\it Total Shuttling Minimized}: Apply only rule~2.

    \item[ ] {\it Minimize Largest Distance}: Apply only rule~1.

    \item[ ] {\it Combined}: Apply both rules; rule~1 takes precedence over rule~2.
\end{enumerate}

All three strategies arise from the basic observation that shuttling is noisy and thus we want to reduce the amount of shuttling. As can be  seen in Fig.~\ref{fig:comparison_main}, the {\it Combined} strategy performs nearly identical to the {\it Minimize Largest Distance} strategy and both significantly outperform the {\it Total Shuttling Minimized} strategy. We give an intuitive explanation for this result. First, since all qubits are shuttled simultaneously and at equal speed (which we assume to be the case, see Sec.\,\ref{subsec:key.assumptions}), the longest shuttling distance determines how long a shuttling phase takes. Reducing the longest shuttling distance reduces the time that all other qubits spend idling and thus serves as a double gain. Second, if the noise caused by shuttling is modeled to grow superlinearly, then reducing the large shuttling distances, even at the cost of increasing some of the shorter shuttling distances, is beneficial for the overall noise. The Gaussian noise model, which we will discuss in Sec.\,\ref{subsec:NoiseModels}, has such superlinear growth and indeed, the difference between strategies is more pronounced for that model compared to other ones. Third, it is clear in the QEC literature that high noise sources such as defective qubits \cite{drouetErrorCorrectionArray2026} have a disproportionately large effect on the resulting logical error rate. This is also the case when a noisy system interacts with an error corrected one, where reducing the error rate linearly reduces the total system error rate exponentially \cite{bultriniBattleCleanDirty2023}. Reducing the largest shuttling distance, rather than the total sum, focuses exactly on eliminating the most defective qubits.

Since strategy {\it Minimize Largest Distance} outperforms strategy {\it Total Shuttling Minimized}, we let rule 1 take precedence over rule 2 once we combine them.

\subsection{Numerical Simulations} \label{ssec:numerics}

We compare our different shuttling strategies and initial layouts by doing memory experiments, since these are well suited to test whether the noise during syndrome extraction is sufficiently controlled by the error correction code.

In a memory experiment, we choose a state of the logical qubits and encode this state in the data qubits according to our chosen error correction code; we repeat the syndrome extraction circuit a number of times to extract the syndrome; then we determine the error that the syndrome suggests occurred using BP-LSD as our decoder. If the correction obtained from the syndrome brings us back into the state representing the correct logical state, then the error correction was successful; otherwise a {\it logical error} occurred. We repeat the memory experiment for up to $10^5$ times to approximate the logical error rate with a resolution of $10^{-5}$. Plotting the logical error rate against the physical error rate on the data qubits informs us on various qualities of this error correction code. We write it as $p_{base}$. See Table \ref{tab:gate_errors} for the ratios of various error sources to $p_{base}$.

We take the ideal circuit of the memory experiment from \cite{QLDPCOrgQLDPC2026}. Since the shuttling noise requires the ability to add customized noise to the circuit, we then build our own code (soon available at \cite{MothquantumSQECRepo}) to add noise to this circuit. The noise model is described in Sec.~\ref{subsec:NoiseModels}. We used stim to simulate the memory experiments.

Our primary goal is to compare different shuttling strategies and noise models to identify important challenges in the practical realization of an error correction code (bivariate bicycle codes, in our case). In addition, we want to compare different shuttling grids and different initial layouts. To do this, we do a parameter sweep over a variety of parameters that we discuss in Sec.\,\ref{methods} and perform a memory experiment for each parameter constellation. Our attention lies on the relative performance of the different shuttling strategies and initial layouts and on the features that remain persistent across the parameter sweep.

\subsubsection*{The Effect of Shuttling Strategies on Logical Error Rates}

\begin{figure}[htpb]
    \includegraphics[width=\columnwidth]{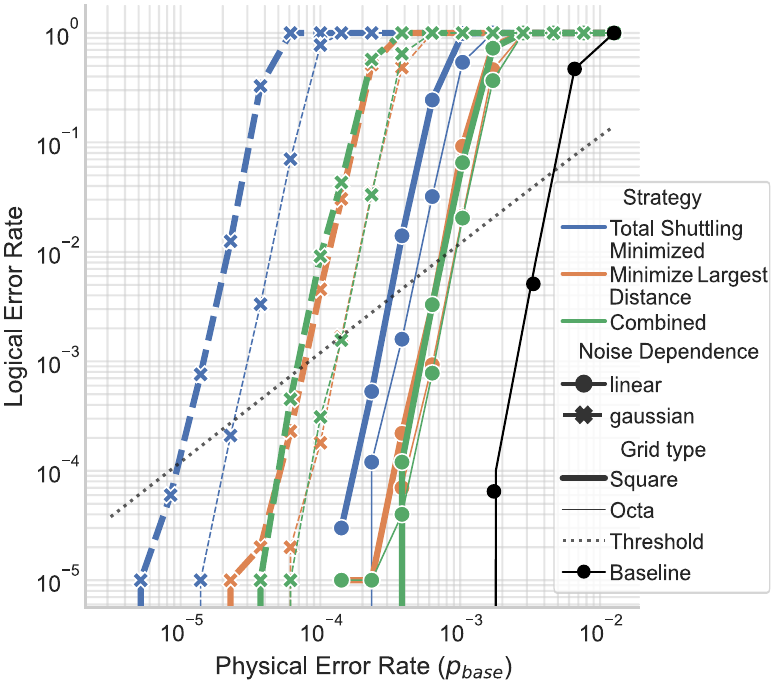}
    \caption{Comparison of different shuttling strategies for various noise models and grids. The Shuttling Error Ratio is set to $0.05$, the Idling to Shuttling Error Ratio is set to $1.8$. State preparation and measurement are set at $2p_{base}$ and the error probability of 2-qubit gates is set at $p_{base}$. Furthermore, the noise is depolarized. Each marker represents the result of a $10^{5}$ shot experiment. We do not show the error bars to keep the plot more legible. Blue lines represent the {\it Total Shuttling Minimized} strategy, which performs the worst; orange lines represent the {\it Minimize Largest Distance} strategy, which alongside the green line representing the {\it Combined} strategy perform better but are comparable. The marker type represents the type of noise model (Gaussian for crosses and circles for linear) described in Sec.~\ref{subsec:NoiseModels}, and the line thickness is the type of connectivity of the shuttling grid (see Fig.~\ref{fig:grids}). The black line with circles represents the baseline, which is only CNOT errors without any shuttling or idling. The dashed black line is the break-even threshold underneath which the logical error rate per qubit is lower than the physical error rate per qubit.}
    \label{fig:comparison_main}
\end{figure}

Throughout our parameter sweep, the strategies {\it Combined} and {\it Minimize Largest Distance} perform  similarly, while strategy {\it Total Shuttling Minimized} performs noticeably worse. The precise difference in performance depends heavily on the noise parameters; for most parameters, strategy {\it Total Shuttling Minimized} performs 1-2 orders of magnitude worse. See figure \ref{fig:comparison_main} for one selected choice of parameters and \ref{fig:comparison_appendix} and \ref{fig:comparison_polarization_appendix} for a broader comparison. We see that parameter choices that increase shuttling noise compared to other noise sources (like 2-qubit gates) increase the difference between the strategies, demonstrating the importance of rule 1 for shuttling noise. Furthermore, we find that this difference can become quite large, exceeding one order of magnitude for fairly mild parameters and reaching two orders of magnitude for parameters that heavily emphasize shuttling noise.

\subsubsection*{Increasing connectivity of the underlying grid can pay off}

The most successful shuttling devices for spin qubits so far are shuttling qubits along 1-dimensional channels \cite{struckSpinEPRpairSeparationConveyormode2024,xueSiSiGeQuBUS2024,volmerImpactLocalValley2026, desmetHighfidelitySinglespinShuttling2025}. We thus model the space in which the qubits can be shuttled as a network of 1-dimensional channels that we call the grid. We use our simulations to compare different grids. We did all memory experiments for two different grids that are discussed in Sec.\,\ref{subsec:Grid} below. The grid with the higher connectivity performs noticeably better, up to two orders of magnitude for noise models that emphasize shuttling noise (see Fig.~\ref{fig:comparison_main}). This informs us about the cost-benefit of building higher connectivity shuttling networks. Doing so comes with new complexities in hardware design and manufacture that can lead to additional sources of errors. Our simulations show that, as long as the added hardware complexity does not cause an enormous increase in noise, the higher connectivity yields a significant improvement in logical error rates.

\subsubsection*{Shuttling Performance is Weakly Dependent on Initial Layout}

\begin{figure}[htpb]
    \includegraphics[width=\columnwidth,trim={0 0 3.3cm 0}, clip]{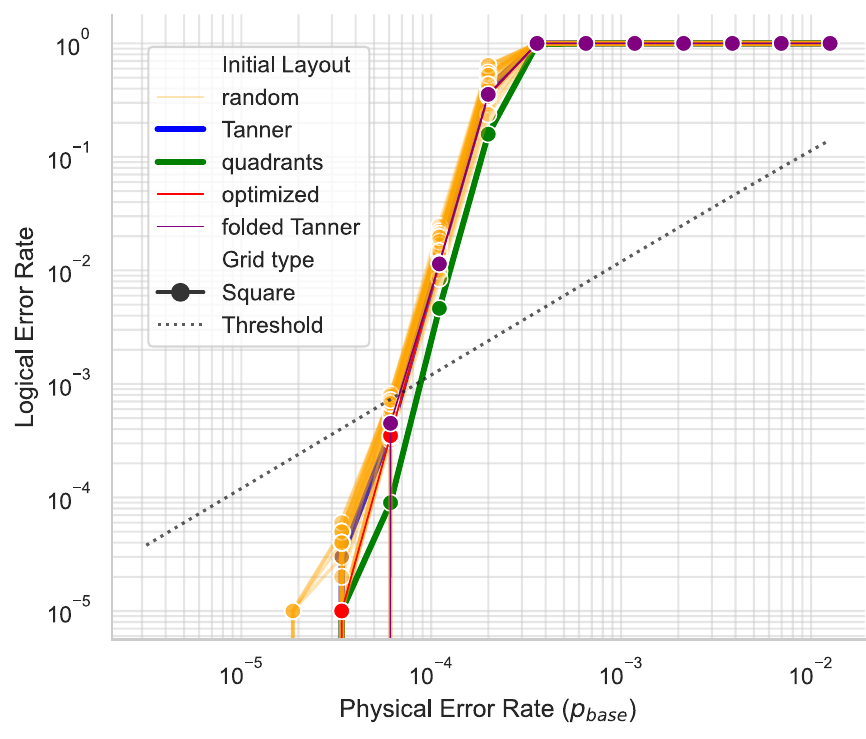}
    \caption{Simulation of a memory experiment for different initial layouts. In orange, we see 25 different random initial layouts, while the other colors display the Tanner, folded Tanner, optimized, and quadrants layout. The data displayed in this figure was obtained by running the memory experiment with the {\it Combined} strategy and a Gaussian noise model.}
    \label{fig:initial_layout_main}
\end{figure}

We did a second datasweep in order to compare the effect of different initial layouts of the qubits. Here too, we compare different strategies, grids, and noise parameters and look for patterns that persist. Testing across 25 random initial layouts and 4 layouts that were chosen with some reasoning in mind (see Sec.\,\ref{ssec:initiallayout}), we find that the initial layout has only little impact on the performance of the error correction. Most notably, the random layouts perform very similarly to the four more carefully chosen layouts. See figure \ref{fig:initial_layout_main} for a comparison across a specific choice of parameters and figure \ref{fig:initial_layout_appendix} for a comparison across various parameter choices. This suggests that there is quite some flexibility in the initial placement of qubits when running the syndrome extraction circuit. This flexibility can be utilized to soften constraints on hardware or circuit optimization.

That said, we frequently find some outliers that perform slightly better than the rest. These outliers usually are among our four non-random initial layouts, but none of them are a consistent outlier.

\section{Methods}\label{methods}

\subsection{Key assumptions} \label{subsec:key.assumptions}

Since there is no existing spin qubit device that could run a large error correction code, we need to make a series of assumptions about what a future device might look like. We state these assumptions, discuss our reasonings for them, and then talk about potential stronger and weaker alternatives.

\begin{enumerate}
    \item The qubits can be moved across a network of $1$-dimensional channels. This network is described as a graph $\Gamma$.

    \item If an edge in $\Gamma$ is undirected, qubits can move along this edge in both directions. In particular, they can move past each other.

    \item If up to two qubits are idling at a vertex, other qubits can move through this vertex. 

    \item The speed at which a qubit moves along an edge is a property of the edge. Formally, each edge has a weight that indicates the time it takes for a qubit to pass through.

    \item Every qubit operation (1-qubit gates, 2-qubit gates, initialization, and measurement) can be executed at any vertex. This is a simplification because the interaction region in hardware may not be at the junctions.

    \item The noise caused by shuttling is the same on every edge.

    \item Qubits are never lost during shuttling, i.e., the transport of the charge is ideal, and all shuttling errors originate from the decoherence of the spin.

\end{enumerate}

Assumption 1 is motivated by the fact that the highest fidelity shuttling devices for spin qubits so far have been $1$-dimensional \cite{struckSpinEPRpairSeparationConveyormode2024,xueSiSiGeQuBUS2024,volmerImpactLocalValley2026, desmetHighfidelitySinglespinShuttling2025}.

Assumption 2 can be achieved by interpreting every undirected edge as two shuttling channels that connect the same pair of vertices but point in opposite directions. Alternatively, T-junctions serve as places where qubits can move past each other. While the first of these two options has not been experimentally shown so far, the swapping of electron patterns via T-junctions has recently been demonstrated at the charge level \cite{beerConveyormodeElectronShuttling2026}. Bus stops adjacent to the shuttling channel, at which a qubit can be parked while other qubits pass, could serve the same purpose \cite{undsethWeightfourParityChecks2026}.

Assumption 3 could be obtained by placing, for example, a `bus stop' that can hold two qubits near every vertex \cite{undsethWeightfourParityChecks2026, kunneSpinBusArchitectureScaling2024}. This allows a qubit-pair to be `stored' at a bus stop while allowing other qubits to pass by. In practice, the timing at which qubits cross vertices or junctions has to be taken into account so that they don't congest a shuttling channel or junction. Alternatively, recent experiments have shown that one can implement single qubit and two-qubit gates on moving qubits, with the two-qubit interaction selected by calibrating the shuttling velocity and distance \cite{matsumotoTwoqubitLogicTeleportation2026,desmetSpinQubitOperations2026}. This circumvents the need to store the qubits near the vertex.

Assumption 4 is based on the idea that the shuttling channels operate like a conveyor belt that moves at a constant speed, however the distances, and therefore shuttling times, are different between paths.

Assumption 5 requires that the appropriate controls are available at every vertex. For instance, one could envision dedicated regions near the vertices that are equipped with micromagnets, enabling single-qubit rotations via electric-dipole spin resonance as well as high-fidelity two-qubit gates such as CZ gates \cite{xueQuantumLogicSpin2022}. Alternatively, gates could be implemented directly on mobile qubits, as discussed above in the discussion of Assumption 3 \cite{matsumotoTwoqubitLogicTeleportation2026,desmetSpinQubitOperations2026}.

Assumption 6 is an idealization. It can be circumvented by incorporating noise differences of different edges into their weights.

Assumption 7 is motivated by the very high charge transfer fidelities reported for conveyor-mode shuttling, e.g., $F = 100^{+0}_{-9\times 10^{-7}}\,\%$ for the transfer between two shuttling lanes through a T-junction~\cite{beerConveyormodeElectronShuttling2026} (see also Ref.~\cite{xueSiSiGeQuBUS2024}). The relevant shuttling errors are therefore those due to spin decoherence, which is what our noise model captures.

In practice, several of these assumptions could be weakened or strengthened without changing our methods substantially, i.e., our method can be generalized to account for more relaxed or stricter constraints. In particular, for the rest of this paper we assume that all edges in $\Gamma$ are undirected and we assume that all edges move qubits at the same speed (though edges may differ in length). On the other hand, assumption 5 may turn out to be  strong. As long as every qubit operation can be executed in sufficiently many places so that every tick of the `theoretical' quantum circuit can be executed as one `hardware' tick on the hardware, our algorithms only need mild modifications that the interested reader may keep track of. If a `theoretical' tick has to be split into several `hardware' ticks, then one has to first restructure the ticks of the circuit, which goes past the focus of this paper.

\subsection{The shuttling grid $\Gamma$}\label{subsec:Grid}

Our first and most important assumption is that the qubits are shuttled on a graph $\Gamma$. We will refer to this graph as the {\it grid}. The choice of grid has an impact on the performance of any error correction code, as we have seen for the Gross code in Sec.~\ref{ssec:numerics} (see Fig.~\ref{fig:comparison_main}). The choice of grid is also heavily restricted by engineering challenges, as it has to be physically built on a chip.

\begin{figure}[htpb]
    \centering
    \begin{tikzpicture}[scale=0.7]

        \begin{scope}
            \node[font=\bfseries] at (1.5, -0.6) {Square Grid};
            
            \foreach \i in {0,...,3} {
                \draw[thick] (\i, 0) -- (\i, 3);
                \draw[thick] (0, \i) -- (3, \i);
            }
            
            \foreach \x in {0,...,3} {
                \foreach \y in {0,...,3} {
                    \filldraw[fill=white, draw=black, thick] (\x,\y) circle (2.5pt);
                }
            }
        \end{scope}

        \begin{scope}[shift={(5,0)}]
            \node[font=\bfseries] at (1.5, -0.6) {Octagrid};
            
            \foreach \i in {0,...,3} {
                \draw[thick] (\i, 0) -- (\i, 3);
                \draw[thick] (0, \i) -- (3, \i);
            }
            
            \foreach \x in {0,...,2} {
                \foreach \y in {0,...,2} {
                    \draw[thick] (\x, \y) -- (\x+1, \y+1);
                    \draw[thick] (\x, \y+1) -- (\x+1, \y);
                }
            }
            
            \foreach \x in {0,...,3} {
                \foreach \y in {0,...,3} {
                    \filldraw[fill=white, draw=black, thick] (\x,\y) circle (2.5pt);
                }
            }
        \end{scope}

    \end{tikzpicture}
    \caption{The Square Grid and the Octagrid. Qubits and interaction zones live on the vertices. The diagonal edges in the Octagrid have length $\sqrt{2}$.}\label{fig:grids}
\end{figure}
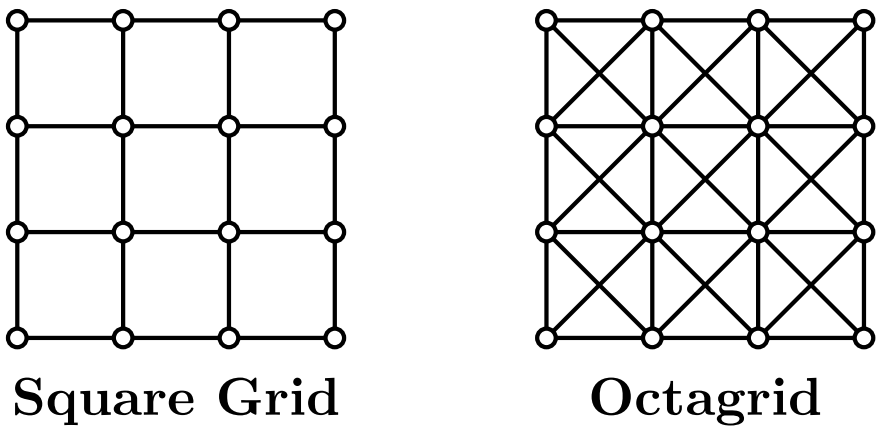

We compare two grids that lend themselves to upscaling (see figure \ref{fig:grids} for mini-examples of both grids). One of the two grids has noticeably higher connectivity. The goal of this comparison is to get a sense of the potential payoff that lies behind the engineering challenges posed by higher connectivity grids. First, we use an $X \times Y$ square grid, where $X$ and $Y$ are suitable parameters for the error correction code in question. \footnote{For the Gross code, this leads us to a $24 \times 12$ square grid.} Second, we consider the {\it octagrid}, by which we mean an $X \times Y$ square grid in which -- for every square -- we have added both diagonal edges. Following the Euclidean geometry of the plane, these diagonal edges have relative length $\sqrt{2}$ compared to the relative length $1$ edges of the square grid.

The planar square grid is a natural extension of current shuttling architectures and has been proposed to host a surface code with shuttling \cite{kunneSpinBusArchitectureScaling2024,yenilenPerformanceSpinQubit2025}. Additional elements that have recently been demonstrated, such as bus stops \cite{undsethWeightfourParityChecks2026} and T-junctions \cite{beerConveyormodeElectronShuttling2026}, make this architecture the most natural one to envision in the near term. However, we also envision that multiple T-junctions and X-junctions, in combination with bus stops, could produce a more complex architecture such as the octagrid. We expect the octagrid to enable more efficient shuttling schedules, given its greater flexibility to connect multiple distant qubits. Here, we aim to quantify how much of an improvement this higher connectivity enables, which will allow us to assess how much is gained by building this possibly more challenging hardware. We find that the octagrid noticeably reduces the logical error rate, except if both the shuttling error ratio and the noise model are maximally noise-minimizing, and by up to almost two orders of magnitude for noise parameters that emphasize shuttling noise (see Figs.~\ref{fig:comparison_main} and~\ref{fig:comparison_appendix}).

Note that we compute the optimized layout in Sec.~\ref{ssec:initiallayout} with respect to the square grid and, for a clean comparison between initial layouts, still use this same layout when running the simulation with the octagrid.

\subsection{Initial Layouts}\label{ssec:initiallayout}

Our shuttling strategies all depend on the initial mapping of the abstract qubits in the syndrome extraction circuit onto the physical qubits moving on the grid. We have tested four different initial qubit layouts and 25 random initial layouts to see how they affect our shuttling strategies. These different layouts all perform quite similarly, see Fig.\,\ref{fig:initial_layout_main} and \ref{fig:initial_layout_appendix}. To understand this, we note that, for a given tick, there are different initial layouts that will lead to the same target coordinates. Therefore, with every shuttling phase, we lose some information about the initial qubit layout. After a few ticks, the current qubit layout has already lost most of the information about the initial qubit layout, leading to similar performance for all of them. This loss of memory of the initial configuration is reminiscent of the mixing behaviour of ergodic dynamics on the space of qubit configurations. Whether this behaviour is generic for qLDPC codes, whose Tanner graphs have a high degree and many long-range edges, or whether it is tied to the translational symmetry of BB codes, which could allow the routing to settle into a periodic sequence of configurations, is an interesting question that we leave for future work.

\begin{figure*}[htpb]
    \centering
    \includegraphics[scale=0.73]{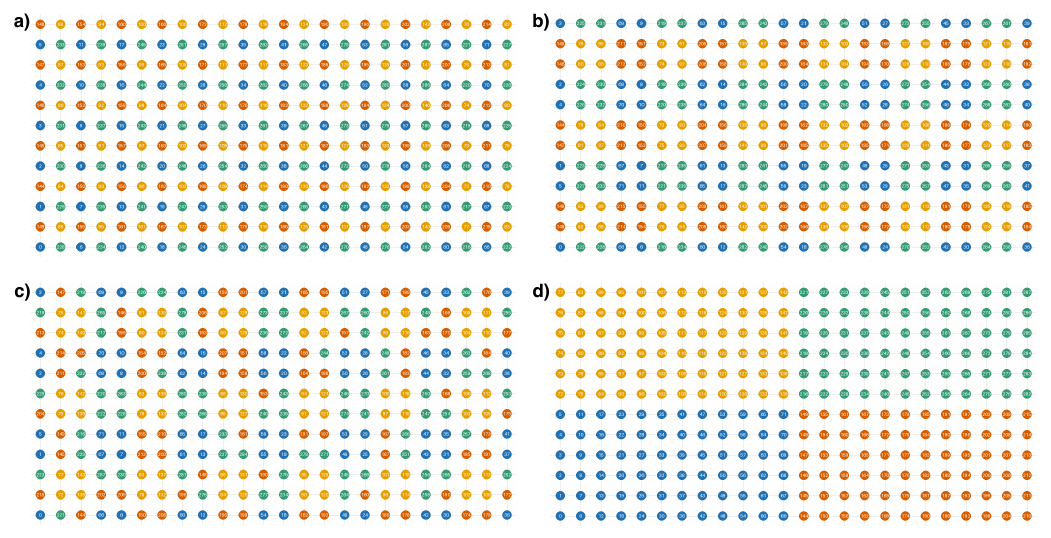}
    \caption{The four different initial layouts. The qubits are color-coded: L-qubits are bLue, R-qubits are oRange, X-checks are red, Z-checks are green. Layout a) is the Tanner layout, b) is the folded Tanner layout, c) is the optimized layout, and d) is the quadrants layout.}
    \label{fig:qubit_layouts}
\end{figure*}

We briefly describe the qubit layouts that we have tested and where they come from. The four non-random layouts were chosen with the expectation that they would perform differently. See figure \ref{fig:qubit_layouts} for a visual display of the first four layouts.

\begin{enumerate}
    \item {\it Tanner}: In this layout, the qubits are placed on the grid in such a way that, for the Gross code, the Tanner graph looks exactly as in standard depictions {\cite[Figure 1 b)]{bravyiHighthresholdLowoverheadFaulttolerant2024}} and can be seen in Fig.~\ref{fig:qubit_layouts}a).

    \item {\it Folded Tanner}: We take the Tanner-layout above and fold it according to {\cite[Figure 2]{pooleArchitectureFastImplementation2025}}, see Fig.~\ref{fig:qubit_layouts}b).

    \item {\it Optimized}: We choose starting coordinates that aim to minimize the largest distance between any two qubits that have to interact at some point during the circuit. Since a fully optimized layout is computationally difficult to obtain, we take the method developed in \cite{pooleArchitectureFastImplementation2025} to produce a reasonable candidate for such an optimized layout. The only difference is that we work with the graph-distance on our grid $\Gamma$, instead of the Euclidean distance. A visualization can be seen in Fig.~\ref{fig:qubit_layouts}c).

    \item {\it Quadrants}: Bivariate Bicycle codes have four different types of qubits: L-qubits, R-qubits, X-checks, and Z-checks. This layout places each type in one quadrant of our grid. We place L-qubits and Z-checks in opposite quadrants. Within each quadrant, the qubits are placed column by column in the order of their numbering. They obtain this numbering from the standard matrix-representation of bivariate-bicycle codes (see \cite{bravyiHighthresholdLowoverheadFaulttolerant2024}). This is shown in Fig.~\ref{fig:qubit_layouts}d). This layout was intended to be a purposefully bad initial layout, but it performs quite well. 

    \item {\it Random}: We chose a random initial layout. We ran the memory experiment for 25 different random initial layouts in order to compare  the performance of a variety of random initial layouts against the four reasoned layouts described above.
\end{enumerate}

As can be seen in Fig.~\ref{fig:initial_layout_main}, the initial qubit layout has little impact on the logical error rate. We thus did the remaining simulations using the optimized initial layout since it follows our rule 1 in minimizing distances between interacting qubits. Fixing the initial layout also reduces the size of our parameter sweep and thus improves computational efficiency.

\subsection{Noise models} \label{subsec:NoiseModels}

The noise models used in this work are approximations to the behaviour of shuttling devices found in the literature. However, different experiments have a large breadth of performance parameters \cite{stanoReviewPerformanceMetrics2022} even when considering the same type of spin qubit. This is also the case when specifically looking at the literature on conveyor mode shuttling \cite{beerConveyormodeElectronShuttling2026,seidlerConveyormodeSingleelectronShuttling2022}. We define our noise model as follows: every gate has an associated depolarizing error channel which is relative to the base physical error rate $p_{base}$, which is also the 2 qubit gate error. State preparation and measurement (SPAM) errors are at $2p_{base}$ and are represented by a bitflip error channel before measurement and a depolarizing error channel after initialization. There are no single qubit gates in our circuits outside of the initial initialization procedure, which is assumed to be noiseless. We assume a shuttling velocity of 10m/s, an edge distance of 400 nm and a two-qubit gate time of 40ns, the same as the time it takes to traverse an edge. This is in line with the literature~\cite{stanoReviewPerformanceMetrics2022} and the edge distance is derived from the SpinBus architecture~\cite{kunneSpinBusArchitectureScaling2024} assuming each vertex has an interaction zone. This information lets us calculate the amount of time qubits are shuttling and idling at any tick given a shuttling route determined by our shuttling strategy. This time is what defines the amount of accumulated error that we then apply to the qubits before a gate is applied.

With this in mind, we perform numerical experiments with various combinations of parameters that are shown in Table~\ref{tab:noise_params} that span from optimistic (at a shuttling ratio of 0.005 per edge) to pessimistic (at a shuttling ratio of 0.1 per edge). In this context the per edge probability refers to the probability of an error when shuttling across one edge. The idling noise has been shown to be somewhat worse than shuttling noise \cite{krzywdaCoherenceProtectionMobile2026}, so we have made our simulations with shuttling and idling noise accumulation over time both equal (1) and with idling being worse than shuttling (1.8), which correspond to the coherence times stated in \cite{krzywdaCoherenceProtectionMobile2026}. 

\begin{table}[htbp]
  \caption{Table showing the different noise parameters. Our experiments can be any combination of one parameter choice per column. Noise accumulation is the function that determines the error probability over the number of edges crossed per qubit for a shuttling step. State preparation and Measurement noise (SPAM) and two qubit (2Q) gate error are toggled on and off to isolate the effect of the shuttling strategies and are linked to $p_{base}$. The shuttling ratio is with respect to $p_{base}$ per edge, and the idling ratio is with respect to the shuttling ratio per edge. Polarization refers to whether we use a polarized noise model~\cite{moncySurfaceCodeThresholdsQubit2026} or a fully depolarizing noise model.}
  \label{tab:noise_params}
  \centering
  \small 
  \setlength{\tabcolsep}{3pt} 
  
  \begin{tabularx}{\columnwidth}{YYYYY}
    \toprule
    Noise accumulation & SPAM and 2Q error  &Shuttling ratio $S_R$& Idling ratio $I_R$& Polarized \\
    \midrule
    Linear      & On  & 0.005 & 1   & On \\
    Exponential & Off & 0.01  & 1.8 & Off\\
    Gaussian    &     & 0.05  &     &    \\
                &     & 0.1   &     &    \\
    \bottomrule
  \end{tabularx}
\end{table}

In the simulations, the shuttling and idling noise is applied as a noise channel preceding a gate, while the gate noise is applied after the action a gate. The equations governing the noise accumulation for shuttling across a distance $d$ are 
\begin{align}
    p_{\text{Gaussian}}^{}(d)&=1-(1-p_{\text{edge}})^{d^{2}},\\
    p_{\text{Exponential}}^{}(d)&=1-(1-p_{\text{edge}})^{d},\\
    p_{\text{Linear}}^{}(d)&= d \cdot p_{\text{edge}}, 
\end{align}
where $p_{\text{edge}}^{\text{shuttling}}=S_Rp_{base}$ with $S_R$ being the shuttling ratio. If a qubit idles for the time it takes to shuttle across a distance $d$, then the noise accumulation is governed by the same equations, except that we set $p_{\text{edge}}^{\text{idling}}=S_RI_Rp_{base}$, with $I_R$ being the idling ratio. Since we assume shuttling noise to be independent from idling noise, the two are multiplied as independent probabilities in the simulation. During every time step where a qubit is neither being shuttled, nor involved in a gate, it accumulates idling noise.

The three equations for noise accumulation above give us the three distinct noise models. We do our comparisons of shuttling strategies, initial layouts, and grids with each of these noise models. They capture different physical regimes of shuttling-induced dephasing. A Gaussian decay corresponds to the standard quasi-static dephasing of a stationary qubit, characterized by $T_2^*$, and also describes a qubit that moves slowly enough for the noise it experiences to remain quasi-static. An exponential decay emerges when the qubit experiences motional narrowing during shuttling \cite{langrockBlueprintScalableSpin2023,struckSpinEPRpairSeparationConveyormode2024,krzywdaCoherenceProtectionMobile2026, boscoHighFidelitySpinQubit2024}, and a linear accumulation corresponds to an exponential decay with a long time scale, e.g., enabled by dynamical decoupling during transport \cite{krzywdaCoherenceProtectionMobile2026,nguyenSuppressingSpinQubit2026, boscoHighFidelitySpinQubit2024}.

\begin{table}
\centering
\caption{Table showing the gate error probabilities relationship to the base physical error rate $p_{\text{base}}$. The entries for shuttling and idling errors refer to the values in Table~\ref{tab:noise_params}, with $S_R$ being the shuttling ratio and $I_R$ being the idling ratio.}
  \label{tab:gate_errors}
  \small 
  \setlength{\tabcolsep}{3pt} 
  
  \begin{tabularx}{\columnwidth}{YY}
    \toprule
    Operation & Error probability \\
    \midrule
    1Q gate        & $\frac{p_{\text{base}}}{10}$   \\
    2Q gate        & $p_{\text{base}}$  \\
    Reset          & $2p_{\text{base}}$      \\
    Measurement    & $2p_{\text{base}}$   \\
    Shuttling      & $\frac{S_Rp_{\text{base}}}{\text{edge}}$\\
    Idling         & $\frac{I_RS_Rp_{\text{base}}}{\text{edge}}$\\
    \bottomrule
  \end{tabularx}
\end{table}

\subsection{The shuttling algorithms}\label{algorithm}

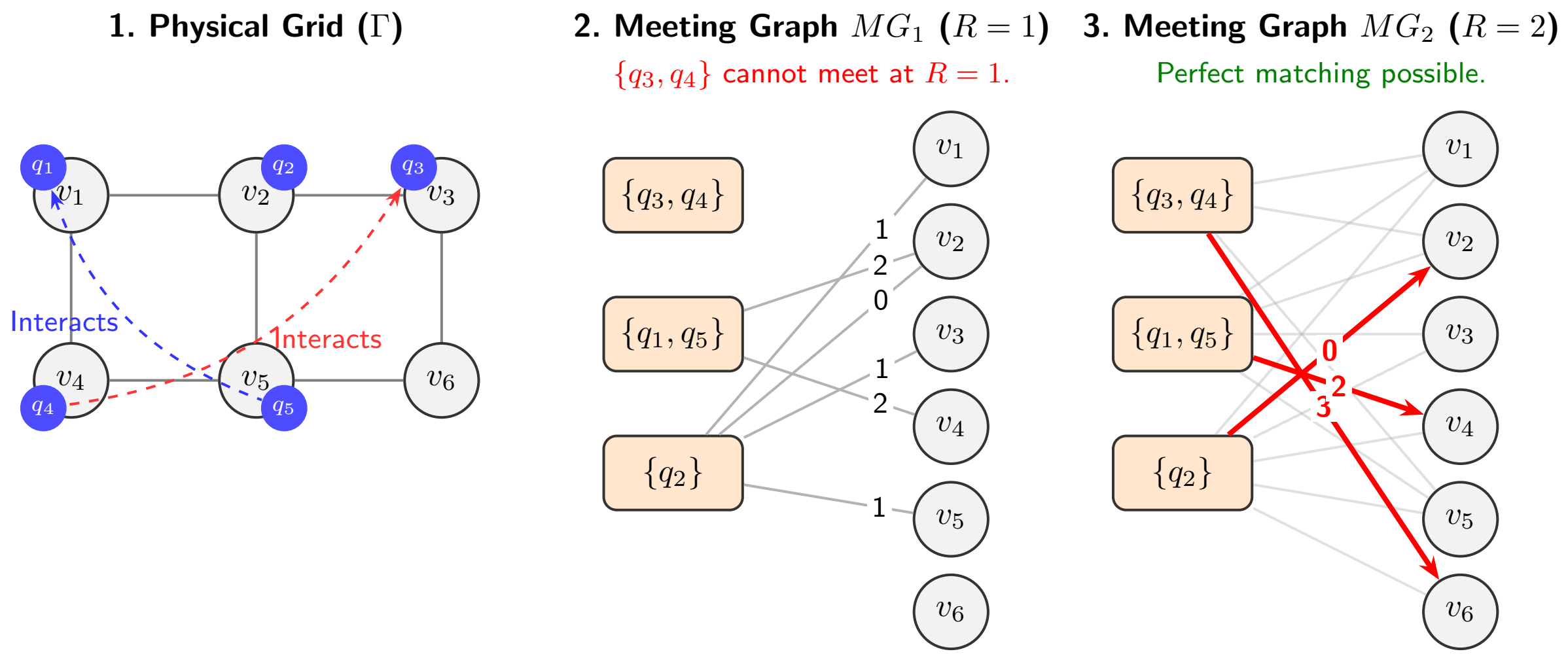
\begin{figure*}[htpb]
    \centering
    \resizebox{\textwidth}{!}{%
    \begin{tikzpicture}[
        gridnode/.style={circle, draw=black!80, thick, fill=gray!10, minimum size=8mm, inner sep=0pt, font=\sffamily},
        qubit/.style={circle, fill=blue!70, text=white, inner sep=1pt, minimum size=5mm, font=\sffamily\footnotesize},
        qset/.style={rectangle, rounded corners, draw=black!80, thick, fill=orange!20, minimum height=8mm, minimum width=15mm, font=\sffamily},
        edge/.style={thick, draw=black!50},
        bipedge/.style={thick, draw=black!30},
        match/.style={ultra thick, draw=red, -{Stealth[length=2.5mm]}},
        weight/.style={fill=white, inner sep=1.5pt, font=\sffamily\scriptsize, text=black}
    ]

    \begin{scope}[shift={(0,0)}]
        \node[font=\sffamily\bfseries] at (2, 3.8) {1. Physical Grid ($\Gamma$)};
        
        \node[gridnode] (v1) at (0,2) {$v_1$};
        \node[gridnode] (v2) at (2,2) {$v_2$};
        \node[gridnode] (v3) at (4,2) {$v_3$};
        \node[gridnode] (v4) at (0,0) {$v_4$};
        \node[gridnode] (v5) at (2,0) {$v_5$};
        \node[gridnode] (v6) at (4,0) {$v_6$};
        
        \draw[edge] (v1) -- (v2) -- (v3);
        \draw[edge] (v4) -- (v5) -- (v6);
        \draw[edge] (v1) -- (v4);
        \draw[edge] (v2) -- (v5);
        \draw[edge] (v3) -- (v6);
        
        \node[qubit] (q1) at ([shift={(-0.3,0.3)}]v1.center) {$q_1$};
        \node[qubit] (q2) at ([shift={(0.3,0.3)}]v2.center) {$q_2$};
        \node[qubit] (q3) at ([shift={(-0.3,0.3)}]v3.center) {$q_3$};
        \node[qubit] (q4) at ([shift={(-0.3,-0.3)}]v4.center) {$q_4$};
        \node[qubit] (q5) at ([shift={(0.3,-0.3)}]v5.center) {$q_5$};

        \draw[thick, dashed, blue!80, -{Stealth[length=2mm]}, {Stealth[length=2mm]}-] 
            (q1) to[bend right=25] node[left, font=\sffamily\scriptsize, text=blue!80] {Interacts} (q5);
        \draw[thick, dashed, red!80, -{Stealth[length=2mm]}, {Stealth[length=2mm]}-] 
            (q3) to[bend left=25] node[right, font=\sffamily\scriptsize, text=red!80] {Interacts} (q4);
    \end{scope}

    \begin{scope}[shift={(6.5,0)}]
        \node[font=\sffamily\bfseries] at (1.5, 3.8) {2. Meeting Graph $MG_1$ ($R=1$)};
        \node[font=\sffamily\scriptsize, text=red] at (1.5, 3.3) {$\{q_3, q_4\}$ cannot meet at $R=1$.};

        \node[qset] (w1_1) at (0, 2) {$\{q_3, q_4\}$};
        \node[qset] (w2_1) at (0, 0.5) {$\{q_1, q_5\}$};
        \node[qset] (w3_1) at (0, -1) {$\{q_2\}$};

        \node[gridnode] (gv1_1) at (3, 2.5) {$v_1$};
        \node[gridnode] (gv2_1) at (3, 1.5) {$v_2$};
        \node[gridnode] (gv3_1) at (3, 0.5) {$v_3$};
        \node[gridnode] (gv4_1) at (3, -0.5) {$v_4$};
        \node[gridnode] (gv5_1) at (3, -1.5) {$v_5$};
        \node[gridnode] (gv6_1) at (3, -2.5) {$v_6$};

        % {q1, q5} -> v2, v4
        \draw[bipedge] (w2_1) -- node[weight, pos=0.8] {2} (gv2_1);
        \draw[bipedge] (w2_1) -- node[weight, pos=0.8] {2} (gv4_1);
        
        % {q2} -> v1, v2, v3, v5
        \draw[bipedge] (w3_1) -- node[weight, pos=0.8] {1} (gv1_1);
        \draw[bipedge] (w3_1) -- node[weight, pos=0.8] {0} (gv2_1);
        \draw[bipedge] (w3_1) -- node[weight, pos=0.8] {1} (gv3_1);
        \draw[bipedge] (w3_1) -- node[weight, pos=0.8] {1} (gv5_1);

    \end{scope}

    \begin{scope}[shift={(12,0)}]
        \node[font=\sffamily\bfseries] at (1.5, 3.8) {3. Meeting Graph $MG_2$ ($R=2$)};
        \node[font=\sffamily\scriptsize, text=green!50!black] at (1.5, 3.3) {Perfect matching possible.};

        \node[qset] (w1_2) at (0, 2) {$\{q_3, q_4\}$};
        \node[qset] (w2_2) at (0, 0.5) {$\{q_1, q_5\}$};
        \node[qset] (w3_2) at (0, -1) {$\{q_2\}$};

        \node[gridnode] (gv1_2) at (3, 2.5) {$v_1$};
        \node[gridnode] (gv2_2) at (3, 1.5) {$v_2$};
        \node[gridnode] (gv3_2) at (3, 0.5) {$v_3$};
        \node[gridnode] (gv4_2) at (3, -0.5) {$v_4$};
        \node[gridnode] (gv5_2) at (3, -1.5) {$v_5$};
        \node[gridnode] (gv6_2) at (3, -2.5) {$v_6$};

        \draw[bipedge, opacity=0.4] (w1_2) -- (gv1_2);
        \draw[bipedge, opacity=0.4] (w1_2) -- (gv2_2);
        \draw[bipedge, opacity=0.4] (w1_2) -- (gv5_2);
        
        \draw[bipedge, opacity=0.4] (w2_2) -- (gv1_2);
        \draw[bipedge, opacity=0.4] (w2_2) -- (gv2_2);
        \draw[bipedge, opacity=0.4] (w2_2) -- (gv5_2);
        \draw[bipedge, opacity=0.4] (w2_2) -- (gv3_2);
        
        \draw[bipedge, opacity=0.4] (w3_2) -- (gv1_2);
        \draw[bipedge, opacity=0.4] (w3_2) -- (gv3_2);
        \draw[bipedge, opacity=0.4] (w3_2) -- (gv4_2);
        \draw[bipedge, opacity=0.4] (w3_2) -- (gv5_2);
        \draw[bipedge, opacity=0.4] (w3_2) -- (gv6_2);

        \draw[match] (w1_2) -- node[weight, pos=0.5, text=red, font=\sffamily\scriptsize\bfseries] {3} (gv6_2);
        
        \draw[match] (w2_2) -- node[weight, pos=0.5, text=red, font=\sffamily\scriptsize\bfseries] {2} (gv4_2);
        
        \draw[match] (w3_2) -- node[weight, pos=0.5, text=red, font=\sffamily\scriptsize\bfseries] {0} (gv2_2);
    \end{scope}

    \end{tikzpicture}
    }
    \caption{A toy example illustrating the shuttling routing algorithm for a system where there are planned interactions between $q_1$ and $q_5$, and $q_3$ and $q_4$ on a $3\times2$ square grid. At $R=1$, the meeting graph $MG_1$ admits no perfect matching covering $V_Q$ because the 1-neighborhoods of $v_3$ and $v_4$ do not intersect. Upon expanding to $R=2$, the algorithm correctly constructs $MG_2$ where $\{q_3, q_4\}$ connects to $v_1, v_2, v_5, v_6$. The red arrows represent a resulting minimum-weight perfect matching with a total routing cost of $5$. Weights for the respective edges are written on the edges and represent the sum of the graph distance traveled by the qubit(s).}
    \label{fig:shuttling_toy_example}
\end{figure*}

All three shuttling strategies use a similar algorithm to compute their respective shuttling routes. For each strategy, we transform their respective optimization problem into finding a minimum weight perfect matching on a (weighted) graph. Finding such a matching is then achieved using standard methods. Figure \ref{fig:shuttling_toy_example} provides a toy example of the algorithm we present here. It may serve as a useful guide and visual point of reference.

Let $T$ be a tick and let $coord$ be the function that associates to every qubit the vertex in $\Gamma$ where it is currently located. Fix an integer $R \geq 0$. We build a bipartite graph $MG_R$ that we call the {\it meeting graph for radius $R$}; see Algorithm \ref{alg:build_meeting_graph} for an algorithmic write-up of the construction. The vertex set of $MG_R$ can be split into two sets: $V_{\Gamma}$ and $V_Q$. The set $V_{\Gamma}$ is the set of vertices of $\Gamma$. The set $V_Q$ is defined as follows: For every qubit $q$, we add the qubit $q$ as a vertex, if it does not participate in a 2-qubit gate during $T$. If $q$ participates in a 2-qubit gate with a qubit $q'$ during $T$, we add the set $\{q,q'\}$ as a vertex. Note that we only add this set once; there is no duplicate vertex induced by the other qubit in the interacting pair.

We now add edges as follows: For every $v \in V_{\Gamma}$ and every $w \in V_Q$, we connect $v$ and $w$ with an edge if and only if for every qubit $q$ in $w$, the distance between $coord(q)$ and $v$ is at most $R$. In other words, we connect $w$ with every vertex in $\Gamma$ that can be reached by every qubit in $w$ by shuttling across a distance of at most $R$.

Our goal is to give every qubit a location to be at for the tick $T$. Qubits should be sent to the same location if and only if they interact via a 2-qubit gate during $T$. Every edge $(v,w)$ in $MG_R$ corresponds to choosing the location $v$ for the qubits in $w$. If $MG_R$ admits a perfect matching (a choice of edges so that every vertex in $V_Q$ is incident to exactly one chosen edge), then we can find such locations for all qubits while having every qubit traveling for a distance of at most $R$.

We arrive at a first algorithm (cf.\,Algorithm \ref{alg:strategy_min_largest}): We start with $R = 0$ and build $MG_R$. We check whether $MG_R$ admits a perfect matching. If not, we increase $R$ by $1$ and build the new graph $MG_R$. We continue until $MG_R$ admits a perfect matching. The perfect matching tells us where every qubit should go. This terminates, since there has to be a perfect matching when $R = \mathrm{diam}(\Gamma)$. (We assume that the number of vertices in $\Gamma$ is at least the number of qubits. Otherwise, there is not enough space on the graph for all qubits during some ticks.)

By construction, this algorithm obeys rule 1 and thus executes the strategy {\it Minimize Largest Distance}. In order to have it obey rule 2 as well, we equip the edges of $MG_R$ with weights. Specifically, the edge $(v,w)$ receives the weight
\[ weight((v,w)) := \sum_{q \in w} d(coord(q),v), \]
where $d$ is the shortest-path-distance in $\Gamma$. In words, the edge $(v,w)$ receives as its weight the total distance needed to move all qubits in $w$ to the location $v$. Instead of searching just for a perfect matching, we now search for a perfect matching with minimum weight. This gives us a new algorithm, Algorithm \ref{alg:strategy_combined}, which optimizes by both rules with rule 1 superseding rule 2. We thus have an algorithm for the strategy {\it Combined}.

To build an algorithm for the strategy {\it Total Shuttling Minimized}, we simply start with $R = \mathrm{diam}(\Gamma)$. For this $R$ the graph $MG_R$ is a complete bipartite graph. Finding a minimum weight perfect matching for this graph optimizes for rule 2. This is Algorithm \ref{alg:strategy_total_min}, which realizes the strategy {\it Total Shuttling Minimized}.

\subsection{Complexity analysis}

Assuming that distance computations on $\Gamma$ can be done in $O(1)$ (which holds for both grids considered here, since their graph distances can be computed in closed form), the complexity of all three algorithms is dominated by the complexity of finding a minimum weight perfect matching. To find this matching, we use the method \texttt{minimum\_weight\_full\_matching} algorithm of the package networkx \cite{hagbergExploringNetworkStructure2008}. Its runtime is $O(\vert V(\Gamma) \vert^3)$. Since we have to repeat this algorithm up to $\mathrm{diam}(\Gamma)$  times, the runtime of our shuttling route computations is dominated by a term of $O\left( \vert V(\Gamma) \vert^3 \cdot \mathrm{diam}(\Gamma) \right)$.

Suppose now that we do not have efficient distance computations on $\Gamma$. In the worst case, we have to use Dijkstra's algorithm to compute distances. To avoid repeat computations, we may cache all distance computations. Doing so leads to a complexity contribution of $\#\{ \text{runs of Dijkstra required} \} \cdot \text{complexity of Dijkstra's algorithm}$. Noticing that the number of edges in $\Gamma$ is proportional to the number of vertices, we may reduce the complexity of Dijkstra's algorithm to $\Theta(\vert V(\Gamma) \vert \cdot \log(\vert V(\Gamma) \vert ) )$. We now want to understand the number of distances that are needed during our algorithm. During the construction of the meeting graph $MG_R$, we need to test the distance of every qubit (which are distributed across at least $\frac{ \vert V(\Gamma) \vert}{2}$ many locations) from every vertex in $\Gamma$. Thus, we have to run Dijkstra's algorithm for at least $\frac{ \vert V(\Gamma) \vert}{2}$ many vertices. Therefore, the complexity of computing and caching all required distances is $\Theta( \vert V(\Gamma) \vert^2 \log( \vert V(\Gamma) \vert))$.

To compare this to the complexity of the \texttt{minimum\_weight\_full\_matching} algorithm, we observe that a bivariate bicycle code is usually placed on a structure that is approximately a square grid. Thus $\mathrm{diam}(\Gamma)$ is roughly proportional to $\sqrt{ \vert V(\Gamma) \vert }$. Under this approximation, we see that the computation of the necessary distances is less expensive than the computation of the minimum weight perfect matching. Our algorithm thus has a complexity of $O(\vert V(\Gamma)\vert^{3} \mathrm{diam}(\Gamma))$, or approximately $O(\vert V(\Gamma) \vert^{\frac{7}{2}})$. Despite this, the complexity of computing the distances is still large enough that more efficient distance computations have a substantial effect on the runtime of our shuttling route calculations in practice

\section{Discussion}\label{Discussion}

In this paper, we have considered two rules for shuttling: 1) minimizing the largest shuttling distance and 2) minimizing the total sum of shuttling distances. Out of these two rules, we built three shuttling strategies, {\it Total Shuttling Minimized, Minimize Largest Distance} and {\it Combined} which we then compared. Our comparison showed that rule 1 is much better at reducing logical errors than rule 2 and this is persistent when varying surrounding parameters like the noise model. Our first reason to focus on these two rules is that we wanted to make a thorough examination of the significance of rule 1. It is a-priori not clear that rule 1 is more important than rule 2, let alone that it yields as substantial an improvement as our results show. Our second reason is that, aside from look-ahead and some considerations concerning the precise path a qubit takes, there are few other promising options how one could optimize. The main alternative is to minimize the largest idling time during a shuttling phase; a strategy whose implementation and testing is left for future work.

We have focused more on running the memory experiment for a broad range of parameters rather than taking the parameters from a specific device. This is because actual spin qubit devices are not yet at the stage where we could get a full set of parameters from a single device that can host a circuit as large as the one we are considering. For this reason, we focused on the features that persist throughout a broad variety of parameters, rather than trying to imagine a concrete fictional device.

The work closest to ours is Ref. \cite{chadwickCAbLECAREfficientlyScheduling2026}, which schedules the syndrome extraction of several qLDPC codes, including the Gross code, on a tileable architecture inspired by SpinBus \cite{kunneSpinBusArchitectureScaling2024}, with unit cells of $\sim$10~$\mu$m. There, the data qubits are stationary and only the ancilla qubits are shuttled, collisions are avoided by a path-planning algorithm that minimizes the duration of each syndrome extraction round, and the circuits are tailored to the dephasing bias of shuttling noise. Here, instead, all qubits are mobile on a dense grid and can pass each other, which allows us to compare shuttling strategies, grid connectivities, initial layouts and noise-accumulation models. Where the two studies overlap, they are consistent: minimizing the duration of a round is closely related to our rule~1, and our simulations quantify its benefit over minimizing the total shuttling distance (1--2 orders of magnitude in the logical error rate). Moreover, the collision-free schedules of Ref. \cite{chadwickCAbLECAREfficientlyScheduling2026} are on average only 12.5\% longer than ideal schedules that ignore collisions, suggesting that relaxing our assumptions~2 and~3 need not change our conclusions, although congestion may be stronger on our dense grid. At $p_{\text{base}}=10^{-3}$, our shuttling ratios correspond to error probabilities of about $1.3\times10^{-4}$--$2.5\times10^{-3}$ per $10~\mu$m of transport in the exponential model, i.e., comparable to or below the lowest value considered in Ref. \cite{chadwickCAbLECAREfficientlyScheduling2026} ($10^{-3}$ per unit cell), at which the Gross code outperforms the surface code, whereas our idling noise is stronger (an effective $T_2$ of about $0.4$--$0.8$~ms for $S_R=0.05$, compared with $10$~ms there). Bias-tailored circuits, collision constraints and a surface-code baseline on the same grids are natural extensions of our work; conversely, our results on grid connectivity and initial layouts address aspects not covered in Ref. \cite{chadwickCAbLECAREfficientlyScheduling2026}. Another work, Ref. \cite{siegel2026nestultrahighrateqldpccodes}, presents a chip topology called NeST, which promises a simple hardware layout able to route certain families of QEC codes efficiently, our work provides quantitative backing for NeST's synchronization routine. 

The algorithms and code we built provide a good starting point for future investigations. Most notably, this work lays the groundwork for a series of potential works. For example: What strategies can be built if one uses look-ahead in the circuit. Can we compare shuttling strategies with the routing methods, like LightSABRE \cite{zouLightSABRELightweightEnhanced2024} that is done in superconducting qubit devices? What happens if we get rid of assumption 5? And finally, do the findings of this paper persist if we look at other bivariate bicycle codes, or even general qLDPC codes \cite{breuckmannQuantumLowDensityParityCheck2021,chadwickCAbLECAREfficientlyScheduling2026}?

We have introduced a series of algorithms that allow us to execute several shuttling strategies, to implement a noise model that takes concrete shuttling routes into account, and to test these strategies in a memory experiment. The memory experiments demonstrate that the choice of shuttling strategy has a substantial impact on the performance of the bivariate bicycle code, provided that shuttling noise is not insignificantly small compared to other noise sources. We see that minimizing the largest shuttling distance reduces shuttling noise far more than simply minimizing the total shuttling distance, giving a clear priority when determining the shuttling for running a quantum circuit. We also see that increasing the connectivity of the underlying grid reduces the logical error probability even while keeping the grid fairly simple. Finally, we observe that the initial layout of the qubits plays a marginal, yet not fully negligible role when minimizing the logical error rate.

\backmatter

%\bmhead{Supplementary information}

%If we need it

\bmhead{Acknowledgements}

This work was partially funded by NCCR SPIN, a National Centre of Competence in Research funded by the Swiss National Science Foundation (grant number 225153).

\section*{Declarations}

\begin{itemize}
\item Funding: N/A
\item Conflict of interest/Competing interests: Authors declare no competing interests.
\item Ethics approval and consent to participate: Not applicable.
%\item Consent for publication: We give consent for publication.
\item Data availability : The full data that support the findings of this study are available from the corresponding author upon reasonable request, but data to reproduce similar figures will soon be %are 
available at \cite{MothquantumSQECRepo}, this is only because of storage limitations.
\item Materials availability: Not applicable.
\item Code availability: The code that supports the findings of this study will soon be %is 
available in github \cite{MothquantumSQECRepo}.
\item Author contribution: MIM designed the algorithm and wrote a majority of the code and related sections in the manuscript (Results and appendices), but also contributed throughout. MTPN and SB reviewed the manuscript at length and proposed many changes and edited the work especially in the parts concerning spin qubits. JRW contributed to the theoretical analysis and provided critical feedback on the manuscript. DB wrote parts of the code and prepared the HPC numerical simulations, in the manuscript he contributed the introduction, numerical simulation, methods and conclusion sections. 
\end{itemize}

\bibliography{SQEC}
\appendix

\section{Building the noisy circuit} \label{app:NoisyCircuit}
The noisy circuit is constructed from the memory experiment circuit through a series of operations. First, the shuttling routes are computed which determine the shuttling time and idling time per qubit. This is done as described in Sec.~\ref{subsec:NoiseModels}. Then, the circuit is iteratively read instruction by instruction, and a new circuit is constructed concurrently with the rule that the shuttling and idling noise is added before the gate and the gate noise is added after the gate. The shuttling and idling noise is exclusively a single qubit Pauli channel, which is either polarized or depolarized depending on the simulation setting and the gate noise can either be a single- or two-qubit channel depending on the gate being a single- or two-qubit gate. This is done on a tick by tick basis, and if a qubit does not have any gates in a tick, then only idling noise is applied equal to the duration of the longest shuttling time plus the gate duration. The process is illustrated in Fig.~\ref{fig:adding_noise}.

\section{The datasweeps} \label{datasweep}

To execute our datasweeps, we built a method that runs a memory experiment for a list of chosen parameters and saves the results. In order to allow for parallelization, one may run this method for one single set of parameters and run all the parameters one wants to check in parallel. We did two major datasweeps. One to compare different initial layouts and one to compare strategies, noise models, and grids.\\

\subsection{Initial layout datasweep} \label{app:initial_layout_datasweep}

In this datasweep, we compare 29 different initial qubit layouts for various strategies, grids and noise model parameters. To reduce the number of simulations, we reduced the variability of many parameters to the following:
\begin{enumerate}
    \item Strategies: We compared strategies {\it Total Shuttling Minimized}, {\it Minimize Largest Distance}, and {\it Combined}.

    \item Noise Sources: Our noise sources are idling, shuttling, 2-qubit gates, measurements, and initializations. We ran every simulation once with SPAM and 2Q error and once without (cf.\,table \ref{tab:noise_params}).

    \item Shuttling Noise Dependence: We compare the Gaussian and exponential noise dependence for shuttling noise, since the linear model is nearly identical to the exponential one.

    \item Grid: We compare the square grid and octagrid.

    \item De-/polarization: We only run the simulation for depolarized noise, since our other simulations show that this choice does not affect any of the comparisons we are doing.

    \item Shuttling Error Ratio: We fix the ratio of Shuttling Error Noise per edge over two-qubit Gate Noise to be $0.05$. As our other datasweep shows, an increase of this ratio simply causes a gradual increase of the difference between different strategies and noise models, depending on how well they deal with shuttling noise and how strongly they emphasize shuttling noise respectively.
\end{enumerate}

\begin{figure*}[htpb]
    \centering
    \includegraphics[scale=0.3]{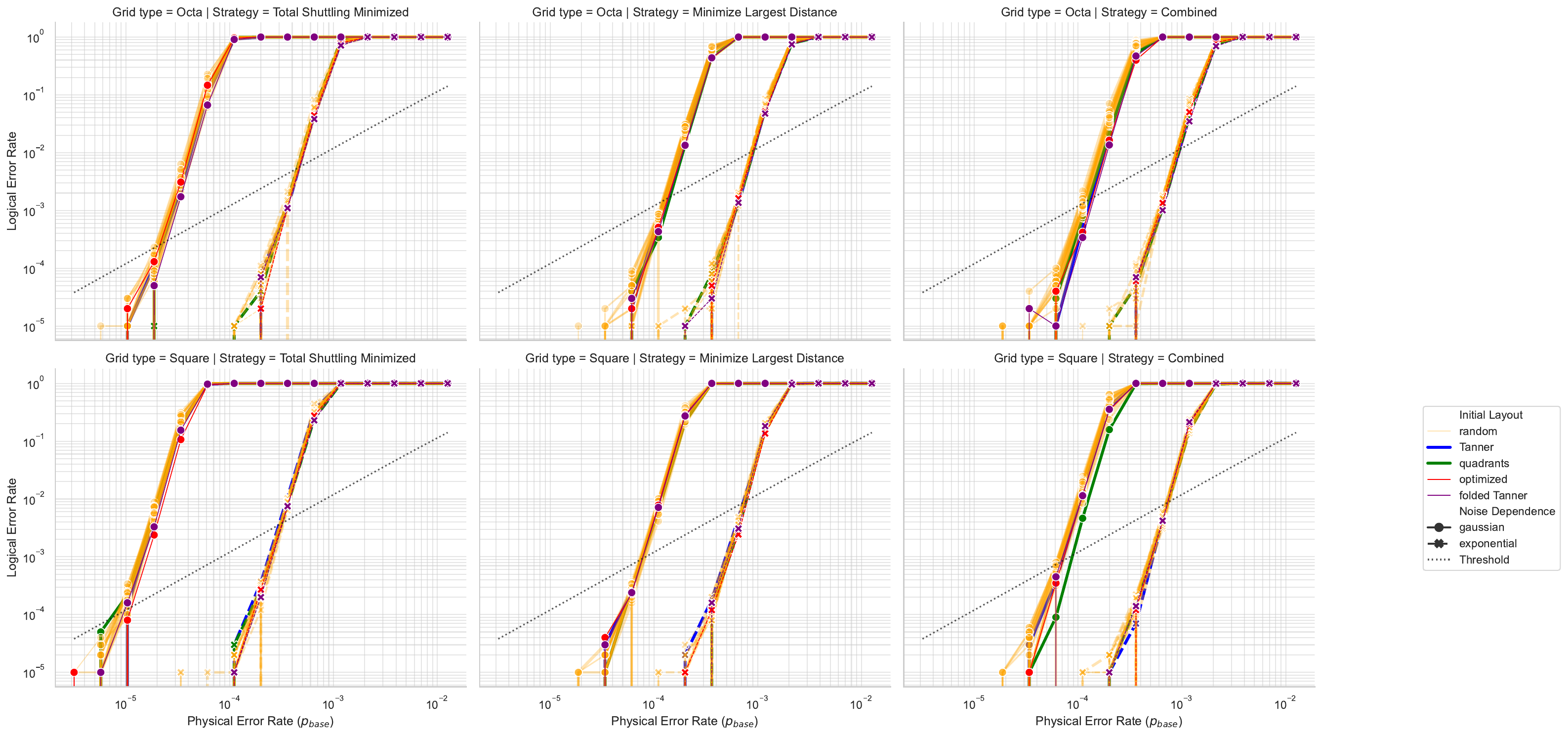}
    \caption{Simulation of a memory experiment for different Initial Layouts. The results shown are for a shuttling ratio $S_R$ of 0.05 and an idling ratio $I_R$ of 1.8, using a depolarized noise model. In addition to the different initial layouts, we compare different shuttling strategies, noise models, and grid types. Colors represent the initial layouts. There are 25 different random layouts which all have the same yellow colour. The dashed lines with cross markers represent exponential noise dependence and solid lines with circular markers represent Gaussian noise dependence. The dashed black line is the threshold line. The columns represent shuttling strategy and the rows represent the grid type. In general, the choice of initial layout does not impact performance consistently.}
    \label{fig:initial_layout_appendix}
\end{figure*}

As we can see from the plots (see figure \ref{fig:initial_layout_appendix}), the initial layout has very limited impact on the performance of the memory experiment. They differ the most if we use Strategy {\it Combined} -- our best strategy -- and a Gaussian Noise Model. As we see in our other datasweep, these are the two choices that amplify the impact of anything shuttling-related the most.

\subsection{Strategy datasweep} \label{app:strategy_datasweep}

The second datasweep aims at comparing different shuttling strategies and different grids for a variety of noise models and noise parameters. The following is a precise list of the parameters we ran the simulations for; these were largely discussed in Sec.~\ref{methods}. 

\begin{enumerate}
    \item Strategies: We test all three strategies

    \item Noise Model: We model the dependence of the shuttling error rate on the shuttling distance in three different ways. By a linear dependence, an exponential dependence, and a Gaussian dependence. 

    \item Grid: For the underlying grid, we compare between the square grid and the octagrid.

    \item De-/polarization: We can adjust the level of polarization in our noise model. We compare only the case of polarization as in \cite{moncySurfaceCodeThresholdsQubit2026} and no polarization.

    \item Initial Layout: Based on our findings in the previous datasweep, we work exclusively with the 'optimized' initial layout for the larger datasweep.

    \item As in the previous datasweep, we run all simulations once with SPAM and 2Q error and once without.

    \item We test four different ratios between the shuttling noise per edge and the two-qubit gate noise. These ratios are $0.005, 0.01, 0.05, 0.1$.

    \item We express the idling noise in terms of the idling noise during the time it takes to shuttle a qubit across one edge. We encode this idling noise in term of its ratio to the shuttling noise per edge. For these idling-to-shuttling ratios, we chose $1, 1.8$. For a justification of the $1.8$ ratio, see \cite{krzywdaCoherenceProtectionMobile2026}.

    \item The physical error ratio that is displayed on the $x$-axis of our plots corresponds to the assumed 2-qubit gate noise. We run the memory experiment for 15 distinct physical error rates ranging from $10^{-4.8}$ to $10^{-1.9}$. They are roughly evenly spaced on a logarithmic scale.

    \item When running syndrome extraction, one has a choice how many rounds of syndrome extraction to run. A common choice is to do as many rounds as the theoretical distance of the error correction code. We followed this principle. In the case of the Gross code, this amounts to $12$ rounds.

\end{enumerate}

\begin{figure*}[htpb]
    \centering
    \includegraphics[scale=0.3]{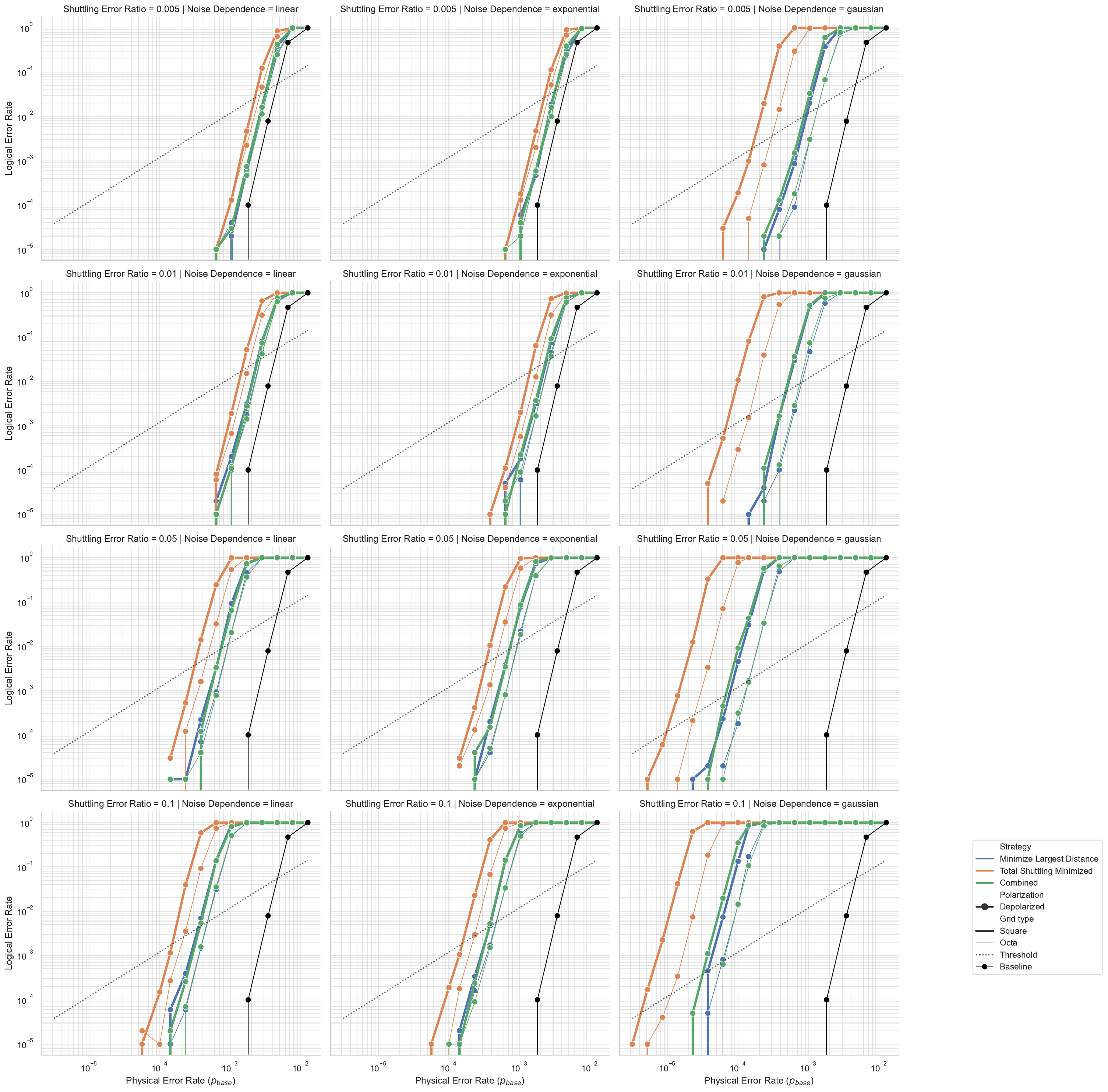}
    \caption{Comparison of different shuttling strategies for various noise models and grids. The Idling to Shuttling Error Ratio $I_R$ is set to $1.8$ using a depolarized noise model. The rows have different shuttling ratios and the columns show different noise dependency. The strategies have different colors and the line thickness represents different grid types. The baseline black line represents a noise model with only CNOT errors and no shuttling or idling noise. This is equivalent to an all to all connectivity and is given as a reference point. The results show that the Octagrid is better than the square grid and that the strategy of minimizing the largest distance or the {\it Combined} strategy are the best.}
    \label{fig:comparison_appendix}
\end{figure*}

\begin{figure*}[htpb]
    \centering
    \includegraphics[scale=0.3]{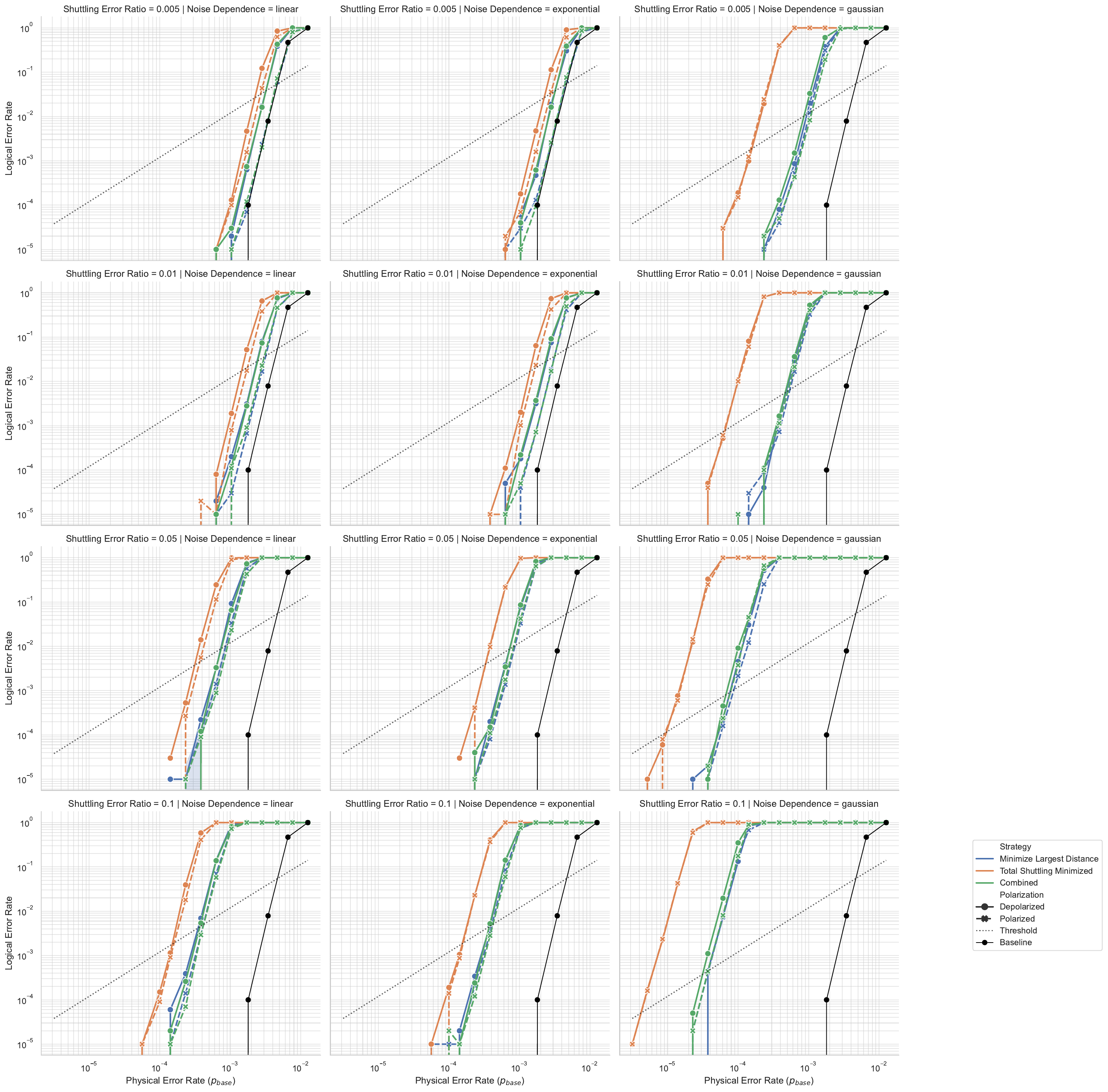}
    \caption{Comparing the effect of polarized and depolarized noise for different strategies, shuttling error rates, and noise dependencies. We see that the polarization makes very little difference in all cases. Just as in figure \ref{fig:comparison_appendix}, the baseline black line represents a noise model with only CNOT errors and no shuttling or idling noise.}
    \label{fig:comparison_polarization_appendix}
\end{figure*}

Figures \ref{fig:comparison_appendix} and \ref{fig:comparison_polarization_appendix} plot a collection of the resulting memory experiments. For practical reasons, we do not display all parameter-constellations in one collection of plots. However, for every choice of parameters, we see the same hierarchy between strategies: Strategy {\it Total Shuttling Minimized} performs significantly worse than the other two, who perform very similarly to each other. We see how the difference in performance gets more pronounced when the noise parameters put more emphasis on shuttling noise. Notably, the difference between the strategies or between the grids becomes more pronounced for larger shuttling error ratios and for the Gaussian noise model.

Furthermore, we see that running the experiment on the octagrid noticeably improves the performance compared to the square grid. Even for small shuttling error ratios, this difference is present and in the most extreme cases, it grows up to almost two orders of magnitude.

\section{Algorithms and figures} \label{app:algorithms}

Here we present the algorithms discussed in this paper to aid in readability and conciseness. Alg.~\ref{alg:build_meeting_graph} is the algorithm used to build the meeting graph which forms the basis of the shuttling strategies. Alg.~\ref{alg:strategy_total_min} represents the algorithm that performs the minimize total shuttling strategy, Alg.~\ref{alg:strategy_min_largest} is the strategy that minimizes longest shuttling distance and Alg.~\ref{alg:strategy_combined} is the algorithm that implements the {\it Combined} strategy. The {\it Combined} strategy consists of the two previous algorithms giving precedence to Alg.~\ref{alg:strategy_min_largest}.

\begin{algorithm}
\caption{Build Meeting Graph}
\label{alg:build_meeting_graph}
\begin{algorithmic}[1]
\Require Graph $\Gamma = (V_\Gamma, E_\Gamma)$, set of qubits $Q$, qubit coordinates $coord$, tick $T$, radius $R$, boolean $\text{useWeights}$
\Ensure Bipartite meeting graph $MG_R = (V_\Gamma \cup V_Q, E_M, weight)$

\State $V_Q \gets \emptyset$
\State $E_M \gets \emptyset$
\State $weight \gets \text{empty map}$

\Statex \Comment{\textbf{Step 1: Construct the qubit vertex set} $V_Q$}
\For{each qubit $q \in Q$}
    \If{$q$ participates in a 2-qubit gate with $q'$ during $T$}
        \State $V_Q \gets V_Q \cup \{\{q, q'\}\}$ \Comment{Sets prevent duplicate inclusion of the pair}
    \Else
        \State $V_Q \gets V_Q \cup \{\{q\}\}$
    \EndIf
\EndFor

\Statex \Comment{\textbf{Step 2: Construct edges and apply weights}}
\For{each $v \in V_\Gamma$}
    \For{each $w \in V_Q$}
        \State $\text{max\_dist} \gets \max_{q \in w} d(coord(q), v)$
        \If{$\text{max\_dist} \leq R$}
            \State $E_M \gets E_M \cup \{(v, w)\}$
            \If{$\text{useWeights}$}
                \State $weight((v, w)) \gets \sum_{q \in w} d(coord(q), v)$
            \EndIf
        \EndIf
    \EndFor
\EndFor

\State \Return $MG_R = (V_\Gamma \cup V_Q, E_M, weight)$
\end{algorithmic}
\end{algorithm}

\begin{algorithm}
\caption{Total Shuttling Minimized}
\label{alg:strategy_total_min}
\begin{algorithmic}[1]
\Require Graph $\Gamma$, set of qubits $Q$, tick $T$
\Ensure A minimum weight perfect matching $M$ assigning target locations
\State $R \gets \mathrm{diam}(\Gamma)$
\State $MG_R \gets \Call{BuildMeetingGraph}{R, \text{True}}$ \Comment{Construct complete distance-weighted graph}
\State $M \gets \Call{MinWeightPerfMatching}{MG_R}$
\State \Return $M$
\end{algorithmic}
\end{algorithm}

\begin{algorithm}
\caption{Minimize Largest Distance}
\label{alg:strategy_min_largest}
\begin{algorithmic}[1]
\Require Graph $\Gamma$, set of qubits $Q$, tick $T$
\Ensure A perfect matching $M$ assigning target locations
\State $R \gets 0$
\State $M \gets \text{null}$
\While{$M = \text{null}$}
    \State $MG_R \gets \Call{BuildMeetingGraph}{R, \text{False}}$ \Comment{Construct unweighted graph}
    \State $M \gets \Call{PerfectMatching}{MG_R}$ \Comment{Standard matching search}
    \If{$M = \text{null}$}
        \State $R \gets R + 1$
    \EndIf
\EndWhile
\State \Return $M$
\end{algorithmic}
\end{algorithm}

\begin{algorithm}
\caption{Combined}
\label{alg:strategy_combined}
\begin{algorithmic}[1]
\Require Graph $\Gamma$, set of qubits $Q$, tick $T$
\Ensure A minimum weight perfect matching $M$ assigning target locations
\State $R \gets 0$
\State $M \gets \text{null}$
\While{$M = \text{null}$}
    \State $MG_R \gets \Call{BuildMeetingGraph}{R, \text{True}}$ \Comment{Construct distance-weighted graph}
    \State $M \gets \Call{MinWeightPerfMatching}{MG_R}$ 
    \If{$M = \text{null}$}
        \State $R \gets R + 1$
    \EndIf
\EndWhile
\State \Return $M$
\end{algorithmic}
\end{algorithm}

\end{document}